\documentstyle[multicol,epsf,aps]{revtex}
\begin{document}
\title{Knots and Random Walks in Vibrated Granular Chains}
\author{E.~Ben-Naim${}^{1,2}$, Z.~A.~Daya${}^{2,3}$,
P.~Vorobieff${}^{2,4}$, and R.~E.~Ecke${}^{2,3}$}
\address{${}^1$Theoretical Division, ${}^2$Center for Nonlinear Studies,
${}^3$Condensed Matter \& Thermal Physics Group} 
\address{Los Alamos National Laboratory, Los Alamos, NM 87545}
\address{${}^4$Department of Mechanical Engineering, 
University of New Mexico, Albuquerque NM 87131}
\maketitle
\begin{abstract}
  
  We study experimentally statistical properties of the opening times
  of knots in vertically vibrated granular chains. Our measurements
  are in good qualitative and quantitative agreement with a
  theoretical model involving three random walks interacting via hard
  core exclusion in one spatial dimension. In particular, the knot
  survival probability follows a universal scaling function which is
  independent of the chain length, with a corresponding diffusive
  characteristic time scale. Both the large-exit-time and the
  small-exit-time tails of the distribution are suppressed
  exponentially, and the corresponding decay coefficients are in
  excellent agreement with the theoretical values.

\end{abstract}
\begin{multicols}{2}
  
  Topological constraints such as knots \cite{kmu} and entanglements
  strongly affect the dynamics of filamentary objects including
  polymers \cite{km,it,srq,bgwb} and DNA molecules \cite{jcw,wc}.
  Typically, large time scales are associated with the relaxation of
  such constraints\cite{dg,de}.  Understanding the physical mechanisms 
  governing the relaxation of such constraints is crucial to
  characterizing flow, deformation, as well as structural properties
  of materials consisting of ensembles of macromolecules, e.g.,
  polymers, gels, and rubber.
  
  Scaling techniques, such as de Gennes-Edwards reptation theory,
  provide a powerful tool for modeling dynamics of topological
  constraints \cite{dg,de}.  These are successful when the precise
  details of the interparticle interactions are secondary relative to
  the geometric effects. However, topological constraints are
  difficult to control experimentally and typically, they can be
  probed only using indirect methods. Here, we introduce a physical
  system where these difficulties are greatly reduced, thereby enabling
  a detailed quantitative comparison with theory.

\begin{figure}
\centerline{\epsfxsize=7cm\epsffile{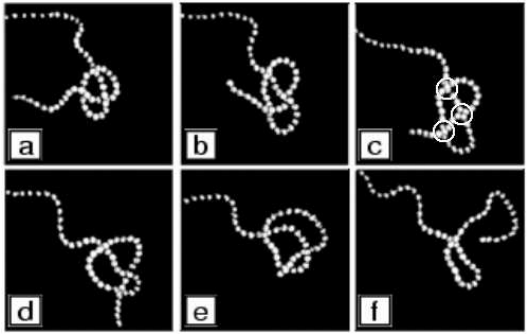}}
\vspace{0.1cm}
\caption{Illustrative snapshots of the vibrated knot experiment,  
taken every ten plate 
oscillation cycles.} 
\end{figure}

In this Letter, we study dynamics of knots in vibrated granular
chains. This system has an appealing simplicity as the ``molecular
weight'' of the chain and the driving conditions can be well
controlled. Additionally, the topological constraints can be
directly observed. We restrict our attention to simple knots and
investigate the time it takes for a knot to open. We find that the
average unknotting time $\tau$ is consistent with a diffusive behavior
$\tau\sim N^2$ where $N$ is the number of beads in the chain.  We also
show that statistical properties of opening times are well described
by a one dimensional model where three random walks, representing the
three exclusion points governing the knot, interact via excluded
volume interactions.  This model provides an excellent approximation
to the knot survival probability.  Furthermore, quantitative
predictions of this model including fluctuations in the exit times, as
well as the coefficients governing the exponential decay of the
extremal tails of the distribution are in excellent agreement with the
measured values.

In the experiments, a simple knot was tightly tied in the center of a
ball chain and placed onto a vibrating plate.  In Fig.~1, we show
images representative of the unknotting process starting from a
tightly knotted chain, Fig.~1a, through intermediate states,
Figs.~1b-e to an unknotted state, Fig.~1f. The chain consists of $N$
hollow nickel-plated steel spheres of diameter $2R_{\rm bead}=2.37\pm
0.02\ mm$ connected by thin rods of diameter $0.52\pm 0.02\ mm$. The
maximum extension between two beads is $0.94\pm 0.01\ mm$, or roughly
$0.8R_{\rm bead}$. The stainless steel plate has a diameter $13.40\  
cm$.  The beads essentially interact via hard core repulsion
(deformation experienced during collisions $< 1\%$), and the
connecting rods act as nonlinear springs.  This bead-spring chain can
be viewed as a ``granular polymer''.  Beads experience dissipative
inelastic collisions with the plate as well as with other beads, and
motion is sustained by injecting energy via the harmonically
oscillating plate \cite{jnb}.

Similar to experiments of vertically-vibrated granular materials
\cite{mus,ou}, given the plate height $z=A\sin \omega t$, with
amplitude $A$, angular velocity $\omega=2\pi \nu$, and frequency
$\nu$, the driving conditions are characterized by the frequency $\nu$
and the dimensionless acceleration $\Gamma=A\omega^2/g$ ($g$ is the
gravitational acceleration).  We examined the parameter range $12
Hz<\nu<16 Hz$ and $1.7<\Gamma<3$, but as statistical properties of the
unknotting time were independent of the driving conditions, we report
results obtained with $\nu=13 Hz$ and $\Gamma=2.40\pm 0.05$ 
(variations in the acceleration across the plate are smaller than 2\%).

\begin{figure}
\centerline{\epsfxsize=7cm\epsffile{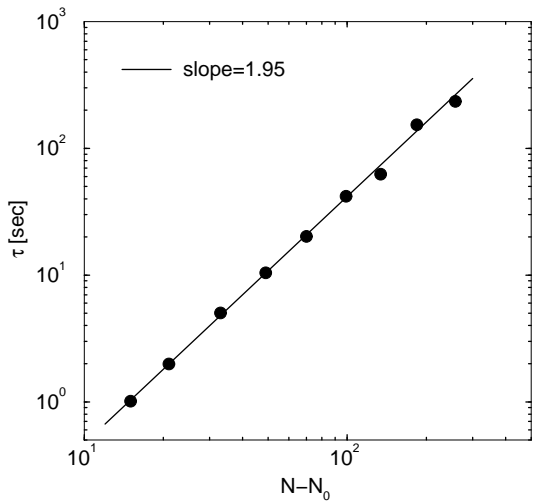}}
\caption{The average unknotting time $\tau$ versus the chain length $N-N_0$.
  Each data point represents an average over 400 measurements.  The
  line represents a power-law best fit.}
\end{figure}

Let us first discuss the dependence of the average unknotting time
$\tau$ on the chain length $N$. We have measured this time for chains
of lengths $30\leq N\leq 273$. The lower limit was dictated by the
knot size as the minimal number of beads associated with the knot is
$N_0=15\pm 1$. Additionally, the error in measurement is significant
for times much smaller than 1 second, and obtaining statistically
significant measurements for large opening times becomes prohibitive.
Nevertheless, the results shown in Fig.~2 provide evidence of
diffusive behavior via the scaling of $\tau$ with $N$:
\begin{equation}
\label{tau}
\tau\sim (N-N_0)^\delta, 
\end{equation}
with $\delta \cong 2.0\pm 0.1$ (the rationale for subtracting the knot
size $N_0$ is given below). A power-law best fit of the data yields
an exponent of $\delta=1.95 \pm 0.07$, and the larger error bar was
obtained by combining possible systematic errors with the statistical
uncertainties.

Next, we examine whether $\tau$ is the only time scale underlying the
distribution of opening times. Let $S(t,N)$ be the probability that a
knot placed on the vibrating plate at time $0$ is still ``alive'' at
time $t$. This probability yields $R(t,N)$, the exit time distribution
$R(t,N)=-{d\over dt}S(t,N)$, and the average time is given by
$\tau=\int dt\,t\, R(t,N)$. The measured survival probabilities
suggest that rather than depending on two parameters $t$ and $N$,
$S(t,N)$ is characterized by a single scaling variable
\begin{equation}
S(t,N)=F(z), \qquad z={t\over \tau}.
\end{equation}
As shown in Fig.~3, apart from systematic deviations for the two 
  shortest chains (this data is not used in further analysis), a
universal scaling function underlies survival probabilities obtained
for different chains.  This scaling behavior is very useful as it
enables us to combine different sets of measurements and therefore
evaluate quantitative characteristics of the distribution with smaller
statistical uncertainties.

\begin{figure}
\centerline{\epsfxsize=6.6cm\epsffile{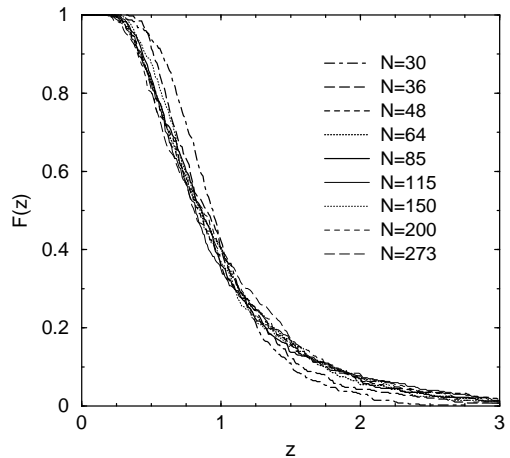}}
\caption{Scaling of the survival probability. The survival
  probabilities corresponding to the data points in Fig.~2 are plotted
  versus the scaling variable $z=t/\tau$.  Although $\tau$ changes by
  two orders of magnitude, the distributions follow a universal
  scaling curve.}
\end{figure}

Our theoretical model is based on simple observations of the knot
dynamics. Whereas the detailed chain motion is complicated, being
determined by wave motion excited in the chain and by collisions
between beads and between beads and the plate, the knot moves via a
series of short range hops of the cross-links constituting it.  When
one of the two external cross links defining the knot crosses one of
the chain ends for the first time, the knot opens.  The diffusive
opening times suggest that the hops are random in direction.

In a chain of length $N$ there are $N-1$ links, and as indicated in
Fig.~1c, the knot motion appears to be subject to three constraints,
namely exclusion points located at positions $1< x_1<x_2<x_3< N-1$.
These exclusion points cannot cross each other and, since the knot
itself contains $N_0$ links, one must have $x_3-x_1>N_0$. Our
theoretical model assumes that (i) these points hop randomly, (ii)
their motion is uncorrelated, {\it i.e.}, all three points hop
independently, and (iii) the $N_0$ links constituting the knot can be
divided equally among the three exclusion points, {\it i.e.},
$x_{i+1}-x_i> N_0/3$.  With these three simplifying assumptions, we
arrive at a model of three identical random walks interacting via
excluded volume interactions on a one dimensional lattice. The finite
$N_0/3$ size of these three ``particles'' merely amounts to an overall
rescaling of the lattice size $N\to N-N_0$.

Within the model framework, the knot survival probability equals the
probability that all three walks remain confined to a finite domain.
We are interested primarily in large chains, and hence, we consider
the continuum limit $x_i\to\infty$, $N\to \infty$ with the variables
$x=x_1/N$, $y=x_2/N$, and $z=x_3/N$ kept finite. Since the three
exclusion points perform independent random walks with identical
diffusivity $D$, the problem is therefore reduced to diffusion in
three dimensions \cite{mef}.  Setting the diffusion coefficient to
unity by redefining the time variable $t\to Dt/N^2$, then
$P(x,y,z,t)$, the probability that at time $t$ the three walks are at
positions $0<x<y<z<1$ respectively evolves according to the diffusion
equation $\partial_t P=\nabla^2 P$.  The initial conditions read
$P(x,y,z,0)=\delta(x-x_0)\delta(y-x_0)\delta(z-x_0)$ with $x_0$ the
knot starting position. The reflecting boundary conditions
$(\partial_x -\partial_y) P\big|_{x=y}=(\partial_y -\partial_z)
P\big|_{y=z}=0$ ensure that the walks do not cross each other.
Finally, the survival probability $S_3(t)$, namely the probability
that all three walks remain confined to within the box boundary, is
obtained by enforcing the absorbing boundary conditions
$P(0,y,z,t)=P(x,y,1,t)=0$, and integrating the probability
$S_3(t)=\int_0^1 dx \int_x^1 dy \int_y^1 dz P(x,y,z,t)$.  The solution 
$P(x,y,z,t)=3!p(x,t)p(y,t)p(z,t)$ can be easily constructed from the
solution of the one dimensional problem $\partial_t
p(x,t)=\partial_{xx}p(x,t)$ subject to the corresponding absorbing
boundary conditions $p(0,t)=p(1,t)=0$, and initial conditions
$p(x,0)=\delta(x-x_0)$. Indeed, the product solution satisfies the
evolution equation as well as the initial and boundary conditions.
Physically, since the walks are identical, the interacting random walk
problem can be mapped to a noninteracting problem by simply exchanging
the identity of the particles whenever their trajectories cross. In
general, $S_m(t)$, the survival probability of $m$ random walks is
\begin{equation}
\label{sm}
S_m(t)=\left [s(t)\right]^m,
\end{equation}
with the single walk survival probability \hbox{$s(t)=\int_0^1 dx\, 
  p(x,t)$} obtained by integrating the well-known solution of the
linear diffusion problem \cite{cy}
\hbox{$p(x,t)=2\sum_{n=1}^{\infty}\sin ({n\pi x_0}) \sin ({n\pi
    x})\exp\big[-(n\pi)^2t\big]$},
\begin{equation}
\label{st}
s(t)={4\over \pi}\sum_{k=0}^{\infty}{\sin [(2k+1)\pi x_0] \over 2k+1}
e^{-(2k+1)^2\pi^2 t}.
\end{equation}
In the following, we set $x_0=1/2$.

Statistical properties of the exit time distribution can be determined
from the survival probability. For example, $R_m(t)$, the exit time
probability distribution function is \hbox{$R_m(t)=-{d\over dt}
  S_m(t)$}, and moments of the exit time distribution are given by
$\langle t^n\rangle_m=\int dt\ t^n R_m(t)$.  As expected, the mean
first passage time decreases as the number of walks increases
$\tau_m\equiv\langle t\rangle_m=1/8,0.073671,0.056213$ for $m=1,2,3$,
respectively.  Rescaling the mean exit times to unity via the scaling
function $F_m(z)=S_m\left(z\langle t\rangle_m\right)$ allows us to
compare the results with the experimental data. As seen from Fig.~4,
the three-random-walk model is in remarkable agreement with the data:
the two distributions agree to within 4\% in the range $z<3$ 
($F>10^{-2}$).  Furthermore, fluctuations in the mean unknotting
times, characterized by the width of the normalized unknotting time
distribution $\sigma^2=\langle z^2\rangle -\langle z\rangle^2=
(\langle t^2\rangle -\langle t\rangle^2)/\langle t\rangle^2$ are
within 1.5\% of each other.  For the experimental data one has
$\sigma=0.62(1)$, whereas the theoretical values are $\sigma=0.81649
(\sqrt{2/3}),0.70495,0.63047$ for $m=1,2,3$.

\begin{figure}
\centerline{\epsfxsize=7cm\epsffile{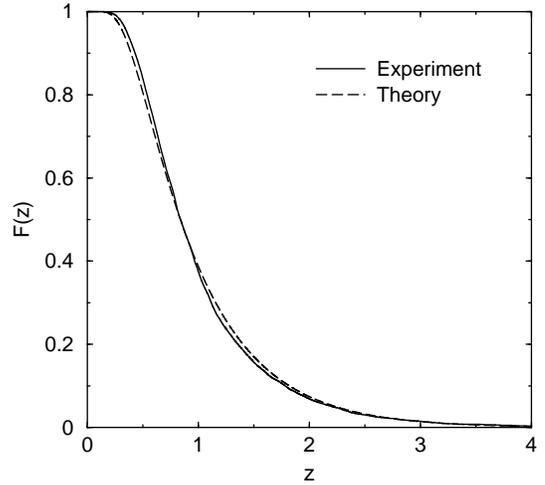}}
\caption{The survival probability $F(z)$ versus the 
  scaling time variable $z=t/\tau$. The experimental curve represents
  6000 data points obtained by aggregating different sets of
  measurements with 100-400 data points in each set. The theoretical
  curve is obtained from
  Eqs.~(\protect{\ref{sm}})-(\protect{\ref{st}}) with $m=3$.}
\end{figure}

Extremal properties can be studied as well.  One may ask ``what is the
probability that the knot opens in a time equal to $k$ times larger or
$1/k$ smaller than the average time?'' (with $k\gg 1$).  The answer to
either questions is ``exponentially small''.  The large tail
statistics follow directly from Eq.~(\ref{st}): the first term in the
series, corresponding to the largest decay time, governs the long time
behavior and $S_m(t)\sim \exp(-m\pi^2 t)$. The large argument tail of
the survival probability is suppressed exponentially
\begin{equation}
\label{large}
F(z)\sim e^{-\beta z}\qquad {\rm for} \qquad z\gg 1,
\end{equation}
with $\beta=m\pi^2 \tau_m$. This exponential behavior is observed
experimentally as shown in Fig.~5. The quantitative agreement is
striking with the experimental value $\beta=1.65(2)$ within $1\%$ of
the theoretical value corresponding to the three random walk model,
$\beta=1.66440$.

In the complementary short time ($t\to 0$) limit, the survival
probability can be found by performing a steepest descent analysis on
the leading large argument ($q\to \infty$) behavior of the exit
probability Laplace transform obtained by differentiating
Eq.~(\ref{st}) $\int dt e^{-qt} \left[-{d\over dt} 
  s(t)\right]=\left[\cosh(\sqrt{q}/2)\right]^{-1}$. In this case one
finds that $1-S_m(t)\sim \sqrt{t}\exp[-1/16t]$, and consequently, the
small argument tail decays exponentially with 
inverse $z$ 
\begin{equation}
\label{small}
1-F(z)\sim \sqrt{z}\,e^{-\alpha/ z} \qquad {\rm for} \qquad z\ll 1,
\end{equation}
where $\alpha=1/16\tau_m$. Here, the theory suggests an additional
algebraic prefactor.  Although the data are consistent with such
behavior as shown in Fig.~5, we can only address the exponential 
behavior with our current statistics.  The difference in this case is
larger: the experimental data yields $\alpha=1.2(1)$ whereas the
theoretical value is $\alpha=1.11184$.  Nevertheless, it is remarkable
that even this more subtle statistic is in good agreement with the
theoretical predictions.

\begin{figure}
\centerline{\epsfxsize=7cm\epsffile{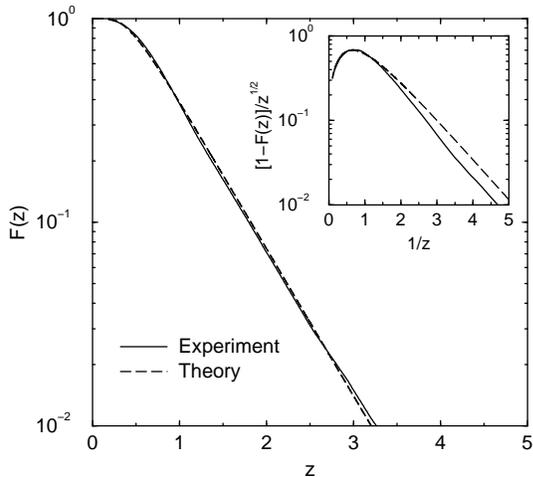}}
\caption{The exponential tails of the survival probability. The experimental
  data of Fig.~4 (solid curves) is plotted versus the three random
  walk theory (dashed curves).  The large argument tail is shown by 
  plotting $F(z)$ versus $z$, while the small argument tail is shown in
  the inset where $\left[1-F(z)\right]/\sqrt{z}$ is plotted versus
  $z^{-1}$.}
\end{figure}

We also examined other predictions of the theory by placing the knot
at off-center starting positions $x_0\ne 1/2$. One can then study
conditional statistics of knots exiting at the near and far ends.
Overall, we find that statistical properties such as the overall
survival probabilities, the conditional survival probabilities, and
the relative fluctuations in the exit times are in reasonable
quantitative agreement with the three random walk model. Additionally,
we verified that the coefficient characterizing the large tail decay
of the distribution is independent of $x_0$ in agreement with
Eqs.~(\ref{sm})-(\ref{st}), reflecting that the initial conditions are
``forgotten'' by long lasting knots.

To test the range of validity of the above results, we studied how
$\tau$ and $F(z)$ depend on the driving conditions.  Fixing the
acceleration at $\Gamma=2.4$, both $\tau$ and $F(z)$ were frequency
independent in the range $12 Hz<\nu<16 Hz$.  Setting the frequency at
$\nu=13 Hz$, $F(z)$ remained the same in the accessible acceleration
range $1.7<\Gamma<3$. This is despite the fact that $\tau$ diverged as
the acceleration approached a critical value $\Gamma_c\cong 1.6$,
below which the cross links did not move and consequently, the knot
remained tied.  These observations are consistent with recent
experiments in vibrated granular layers where the behavior is governed
primarily by $\Gamma$ \cite{mus,ou}.  Additionally, by doubling the
sidewall diameter, and measuring $\tau$, $F(z)$, and $\sigma$, we
confirmed that the effects of the sidewalls were negligible even for
the longest chain. In short, our findings suggest that the driving
parameters ($\Gamma$, $\nu$) and the chain parameters merely determine
the hopping rate $D$, and  that $D$ is uniform along the
chain and independent of the chain length.  In units of the
characteristic time scale, $\tau$, opening time statistics are given
by a universal scaling function $F(z)$.

The most significant assumption made in our model is that the motion
of the exclusion points is uncorrelated. While such correlation is
present for small knots, possibly responsible for the larger
discrepancy in the short exit time tail and in the survival
probability of  short chains, it is possible, however, that it becomes
negligible beyond some fixed correlation length.  In any case, the
random walk model can be useful for characterizing isolated
topological constraints. Measurements of $\tau(N)\simeq
\tau_3(N-N_0)^2/D$ can be used to extract the constraint size $N_0$ as
well as the hopping rate $D$. For instance, minimizing the statistical
uncertainty in the power law best fit to the data in Fig.~2 yields
$N_0\cong 15.2$, consistent the actual knot size $N_0=15\pm 1$, while
the hopping rate $D=11\pm 1\ sec^{-1}$ is found to be comparable with
the frequency $\nu=13Hz$.  Furthermore, the effective number of
constraints $m$ can be deduced by comparing the measured $F(z)$ with
the theoretical $F_m(z)$.

In conclusion, vibrated granular chains provide a useful tool for
probing dynamics of topological constraints.  As the vibrating plate
effectively plays the role of a heat bath, constantly supplying the
system with energy, this system may prove useful for studying issues
of current interest in polymer dynamics as well as in 
granular media.

We thank David Egolf and Zolt\'an Toroczkai for useful discussions.
This research is supported by the US DOE (W-7405-ENG-36)
and by the Canadian NSERC.

\end{multicols}

\begin{thebibliography}{99}
  
\bibitem{kmu}K.~Murasugi, {\it Knot Theory and its Applications}, 
(Birkh\"auser, Boston, 1996).
\bibitem{km}K.~Koniaris and M.~Muthukumar Phys. Rev. Lett. {\bf 66},
  2211 (1991).  
\bibitem{it}K.~Iwata and M.~Tanaka, J. Chem Phys. {\bf
    96}, 4100 (1992).  
\bibitem{srq}S.~R.~Quake, Phys. Rev. Lett. {\bf
    73}, 3317 (1994).  
\bibitem{bgwb}E.~Ben-Naim, G.~S.~Grest, T.~A.~Witten, 
and A.~R.~C.~Baljon, Phys. Rev. E {\bf 53}, 1806 (1996).  
\bibitem{jcw}J.~C.~Wang, J. Mol Bio {\bf 55}, 523 (1971).
\bibitem{wc}S.~A.~Wasserman and N.~R.~Cozzarelli, Science {\bf 232},
  951 (1986).  
\bibitem{dg} P.~G. de Gennes, {\it Scaling Concepts in Polymer Physics} 
(Cornell, Ithaca, 1979).  
\bibitem{de} M.~Doi and S.~F.~Edwards, {\it The Theory of Polymer 
Dynamics} (Clarendon Press, Oxford, 1986).  
\bibitem{jnb}H.~Jaeger and S.~Nagel, R.~P.~Behringer, Rev. Mod.
    Phys. {\bf 68}, 1259 (1996).  
\bibitem{mus}F.~Mello, P.~B.~Umbanhowar, and H.~L.~Swinney, 
Phys. Rev. Lett. {\bf 75}, 3838 (1995).  
\bibitem{ou} J.~S.~Olafsen and J.~S.~Urbach, Phys. Rev.
  Lett.  {\bf 81}, 4369 (1998).  
\bibitem{mef}M.~E.~Fisher, J. Stat  Phys. {\bf 34}, 667 (1984).  
\bibitem{cy}H.~S.~Carslaw and
  J.~C.~Yaeger, {\it Conduction of Heat in Solids}, (Clarendon Press,
  Oxford, 1959).
\end{thebibliography}
\end{document}